\documentclass[10pt]{amsart}
\usepackage{amssymb}
\usepackage{enumerate}
\usepackage{epsf}
\usepackage{psfrag}
\usepackage{graphics}
\usepackage[dvips]{graphicx}
\usepackage{xcolor}
\usepackage{hyperref}
\usepackage{graphicx}
\usepackage{wrapfig}
\usepackage{float}
\DeclareGraphicsExtensions{.eps,.art,.ART,.ps}
\usepackage{mathtools}
\usepackage{graphicx,caption}
\usepackage{subfig}
\newcommand{\bbR}{\mathbb{R}}      

\newcommand{\grad}{\operatorname{grad}}

\begin{document}
\begin{abstract}
We study the motion of an elastic rigid rod which is radially free-falling towards a Schwarzschild black hole. This is accomplished by reducing the corresponding free-boundary PDE problem to a sequence of ODEs, which we integrate numerically. Starting with a rod at rest, we show that it is possible to choose its initial compression profile so that its midpoint falls substantially faster, or slower, than a free-falling particle with the same initial conditions. This seems to be a purely kinematic effect, since on average there is no net transfer of elastic energy to mechanical energy. 
\end{abstract}
%
%
\title{Free-falling motion of an elastic rigid rod towards a Schwarzschild black hole}
\author{Lu\'{i}s Machado, Jos\'{e} Nat\'{a}rio$^*$ and Jorge Drumond Silva}
\thanks{$^*$Corresponding author \href{mailto:jnatar@math.ist.utl.pt}{\tt <jnatar@math.ist.utl.pt>}} 
\address{Center for Mathematical Analysis, Geometry and Dynamical Systems, Mathematics Department, Instituto Superior T\'ecnico, Universidade de Lisboa, Portugal}
%
%
\maketitle
%
%
%
%
\thispagestyle{empty}
%
%
%
\section*{Introduction}
It is well known that the trajectories of free-falling test particles in General Relativity are given by timelike geodesics.  The motion of extended bodies, however, is far more complicated \cite{Dixon70a, Dixon70b, Dixon74}, and the geodesic approximation is only valid in the asymptotic limit when the size of the body is much smaller than the local radius of curvature \cite{GerochJang75, EhlersGeroch04, PPV11}. In fact, extended bodies often exhibit surprising motions (sometimes dubbed ``swimming" or ``gliding" motions), examples of which have been studied in \cite{Wisdom03, GMM06, Harte07, GM07, BDH09, SMV16, MendesPoisson17, VeselyZofka19, Harte20, VeselyZofka21}.

In Classical Mechanics, the simplest model to describe the motion of an extended body is to assume that it is rigid (meaning undeformable). In General Relativity, however, causality considerations prevent the use of such a simplistic assumption. Therefore, relativistic elasticity has become the standard tool to study simple extended bodies \cite{AOS95, ABS09, BW07, BS08, Natario14, BS17, Natario18, JBE23}, as it offers a fully coherent framework within General Relativity which not only models the bulk motion but also accounts for the inevitable deformations that any real object is subject to.

In this work, we focus on the motion of a rigid rod, that is, a $1$-dimensional elastic object whose speed of sound is equal to the speed of light\footnote{Note that ``rigid" is taken to mean ``as rigid as possible", implying the largest admissible speed of sound. In Classical Mechanics this means ``undeformable", with an infinite speed of sound; in General Relativity it implies that the speed of sound is equal to the speed of light.}, falling radially towards a Schwarzschild black hole. This leads to a free boundary problem for the wave equation on the $2$-dimensional Lorentzian manifold corresponding to the radial direction. Using harmonic coordinates (essentially d'Alembert's method for solving the wave equation along null directions), we turn this problem into a sequence of ODEs, which we then integrate numerically for a few illustrative examples, concentrating on the strong field region near the black hole. Specifically, we analyze the motion of a rod starting at rest, for three different choices of the initial compression factor: completely relaxed (for comparison purposes), and two cases where the two halves of the rod are uniformly compressed by different amounts. The soundness of the numerical integration is checked by computing the conserved energy of the rod. We also compute the (non-conserved) elastic energy, whose value seems to oscillate without any discernible trend.

The organization of the paper is as follows: in Section~\ref{section1} we briefly review the general theory of one-dimensional rigid elastic bodies moving in two-dimensional spacetimes. In Section~\ref{section2} we set up the algorithm for solving the equation of motion of a rigid rod moving radially on a general static spherically symmetric spacetime. Section~\ref{section2.5} contains a derivation of the expression for the rod's conserved energy, as well as for its elastic energy. Finally, we numerically integrate the equation of motion for the three examples mentioned above in Section~\ref{section3}. We conclude in section~\ref{section4} by discussing the results and suggesting further research problems.

We follow the conventions of \cite{MTW73}; in particular, we use a geometrized system of units, for which $c=G=1$.
%
%
%
\section{Elastic rod}\label{section1}
We are interested in the motion of an elastic rod, that is, a one-dimensional elastic body, along a fixed direction (namely radial motion in a spherically symmetric spacetime). Therefore, it suffices to consider two-dimensional spacetimes $(M,g)$. The elastic rod can be modeled by a scalar field $\lambda : M \to \mathbb{R}$ with nonvanishing spacelike gradient, whose level sets correspond to the worldlines of the particles that constitute the rod (see \cite{Natario14}). If we choose $\lambda$ to give the proper length along the undeformed rod, then the local compression factor is the positive function $n$ such that
\begin{equation}
n^2 = {|\grad \lambda|}^2 = ({\nabla}^{\alpha}\lambda)({\nabla}_{\alpha}\lambda) \, .
\end{equation}
Since we are considering only one spatial dimension, the energy-momentum tensor of the elastic rod is necessarily that of a perfect fluid, that is,
\begin{equation}
T_{\alpha\beta} = (\rho + p) u_\alpha u_\beta + p g_{\alpha\beta} \, ,
\end{equation}
where $\rho$ is the energy density (per unit lenght), $p$ is the pressure (which is actually a force in this $(1+1)$-dimensional setting) and $u$ is the unit tangent covector to the rod particle's worldlines.

As mentioned in the introduction, we will assume that the rod is rigid, that is, its speed of sound equals the speed of light, meaning that
\begin{equation}
\frac{dp}{d\rho} = 1 \Rightarrow \rho = \rho_0 + p \, ,
\end{equation}
where $\rho_0$ is the (constant) relaxed density. Under this assumption, as shown in \cite{Natario14}, the rod's energy-momentum tensor can be written as
\begin{equation}\label{eq:energymomentumtensor}
T_{\alpha\beta} = \rho_0 \left[ {\nabla}_{\alpha} \lambda \, {\nabla}_{\beta}\lambda - \frac12 ({\nabla}^{\mu}\lambda\,{\nabla}_{\mu}\lambda) g_{\alpha\beta} - \frac12 g_{\alpha\beta} \right] \, ,
\end{equation}
which is, up to a multiplicative constant and a covariantly constant additive term, the energy-momentum tensor for a massless scalar field $\lambda$. Therefore, as is well known, $\lambda$ must satisfy the wave equation:
\begin{equation}\label{eq:waveeq}
    \Box \lambda = 0 \, .
\end{equation}

Generically, the motion of a finite rod leads to a free boundary problem with nonlinear boundary conditions for this equation.\footnote{In this work we will regard the rod as a test body, that is, we ignore its backreaction on the metric. Therefore, we just have to solve the wave equation on a fixed background metric.} Such problems are highly nontrivial, and a careful choice of comoving coordinates, for which the boundary remains fixed, is typically the best approach to study them. In this simple $(1+1)$-dimensional setting, an obvious choice of comoving coordinate is the function $\lambda$ itself, since the free boundary, corresponding to the endpoints of the rod, is given by level sets of this function. As shown in \cite{Natario14}, one can always choose a (conjugate) time function $\tau:M \to \bbR$, also satisfying the wave equation, such that the metric in the (harmonic) coordinates $(\tau,\lambda)$ is written as
\begin{equation} \label{proper}
    ds^2 = \frac{1}{n^2(\tau,\lambda)} \left( - d\tau^2 + d\lambda^2 \right) \, .
\end{equation}
Conversely, if the metric can be written in the form \eqref{proper} then $\lambda$ is a solution of the wave equation. Therefore, a method for solving the problem of the rod's motion is simply to find a harmonic coordinate system.
%
%
\section{Motion on static spherically symmetric spacetimes}\label{section2}

We now focus on the case of radial motion of the rod in a general static spherically symmetric spacetime. The metric on the totally geodesic $2$-dimensional submanifold $M$ corresponding to a fixed radial direction can be written as
\begin{equation}
    ds^2 = -F(r)dt^2+\frac{1}{G(r)}dr^2 \, ,
\end{equation}
where $r$ is the usual areal radius coordinate and $F$ and $G$ are smooth positive functions depending on $r$. Changing to the null coordinates
\begin{equation}\label{eq:nullcoords}
    \begin{cases}
    u=t-x\\
    v=t+x
    \end{cases} \, , \quad \text{where} \quad x(r) = \int \frac{dr}{\sqrt{F(r)G(r)}} \, ,
\end{equation}
the metric of the $2$-dimensional spacetime takes the form
\begin{equation}\label{eq:nullmetric}
    ds^2=-F\left(r\left(\frac{v-u}{2}\right)\right)du \: dv \, ,
\end{equation}
where $r(x)$ is the inverse function of $x(r)$ in equation \eqref{eq:nullcoords}.

On the other hand, changing to null coordinates in the harmonic coordinate system associated to the rod,
\begin{equation}\label{eq:nullcoords2}
    \begin{cases}
    U=\tau-\lambda\\
    V=\tau+\lambda
    \end{cases} \, ,
\end{equation}
the metric \eqref{proper} is represented as
\begin{equation}\label{eq:nullproper}
    ds^2 = -\frac{1}{n^2(U,V)}\,dU\, dV \, .
\end{equation}

Since $(u,v)$ and $(U,V)$ are both future-directed null coordinate systems for our $2$-dimensional spacetime\footnote{We are assuming that the time functions $t$ and $\tau$ induce the same time orientation.}, they can be rescaled into each other, in the sense that there must exist smooth increasing functions $f$ and $g$ such that
\begin{equation}\label{eq:nullcoords3}
    \begin{cases}
    u=f(U)\\
    v=g(V)
    \end{cases} \, .
\end{equation}
Inserting this in equation \eqref{eq:nullmetric} yields
\begin{equation}\label{eq:nullrest}
    ds^2=-F\left(r\left(\frac{g(V)-f(U)}{2}\right)\right)f'(U)\,g'(V) \, dU\, dV \, .
\end{equation}
From \eqref{eq:nullproper} and \eqref{eq:nullrest} we obtain the following equation, which fully determines the motion of the elastic rod:
\begin{equation}\label{eq:motion}
    \frac{1}{n^2(U,V)} = F\left(r\left(\frac{g(V)-f(U)}{2}\right)\right)f'(U)\,g'(V) \, .
\end{equation}
Given initial and boundary conditions, our task is then to determine the functions $f$ and $g$ by solving equation \eqref{eq:motion} in the region corresponding to $\tau>0$ and $0<\lambda<L$, where $L$ is the proper length of the undeformed rod.

As the first initial condition, we will impose that the rod is initially at rest:
\begin{equation}
    \frac{\partial x}{\partial \tau}(0,\lambda)=0 \, , \qquad 0<\lambda<L \, ,
\end{equation}
which is equivalent to
\begin{equation} \label{eq:derivs}
    g'(V)=f'(-V) \, , \qquad 0<V<L \, .
\end{equation}
Integrating this equation, we obtain
\begin{equation}\label{eq:1initial}
    g(V)+f(-V) = 0 \, , \qquad 0<V<L \, ,
\end{equation}
where we set the integration constant to zero by fixing $t=0$ when $\tau=0$. Additionally, we have to impose a second initial condition to define the initial state of the elastic rod: the values of the compression factor at $\tau=0$, that is, $n(-V,V)$ for $0<V<L$.

As boundary conditions, we will impose that the rod is not being pulled or pushed at its ends. This corresponds to setting the values of the compression factor at both ends of the rod equal to one for all time (see \cite{Natario14}), that is, $n(V,V)=1$ for $V>0$ and $n(V-2L,V)=1$ for $V>L$.

Let us suppose that the endpoint of the rod corresponding to $\lambda=0$ is initially at $r=r_0$. Mathematically, this can be expressed by
\begin{equation}
    \frac{g(0)-f(0)}{2}=x(r_0) \, ,
\end{equation}
which, together with equation \eqref{eq:1initial}, implies
\begin{equation} \label{initial}
    \begin{cases}
    g(0)=x(r_0)\\
    f(0)=-x(r_0)
    \end{cases} \, .
\end{equation}

In the segment corresponding to $\tau=0$, the equation of motion \eqref{eq:motion} becomes
\begin{equation}
    \frac{1}{n^2(-V,V)} = F\left(r\left(\frac{g(V)-f(-V)}{2}\right)\right)f'(-V)\,g'(V) \, , \qquad 0<V<L \, .
\end{equation}
Combining this result with equation \eqref{eq:1initial}, we obtain
\begin{equation} \label{eq:gderiv}
    \frac{1}{n^2(-V,V)} = F(r(g(V)))(g'(V))^2 \, , \qquad 0<V<L \, ,
\end{equation}
which, using the  with initial condition \eqref{initial}, can be solved to determine $g(V)$ for $0<V<L$, since the values of $n(-V,V)$ in this interval are known. The values of $f(U)$ for $-L<U<0$, in turn, can be obtained from equation \eqref{eq:1initial}.

Setting $U=V-2L$, the equation of motion  \eqref{eq:motion} becomes
\begin{equation}\label{eq:1}
    \frac{1}{n^2(V-2L,V)} = F\left(r\left(\frac{g(V)-f(V-2L)}{2}\right)\right)f'(V-2L)\,g'(V) \, .
\end{equation}
This equation can be solved to determine $g(V)$ for $L<V<2L$, since the values of $n(V-2L,V)$ in this interval are known, as are the values of $f(U)$ for $-L<U<0$. Similarly, by setting $V=U$, the equation of motion  \eqref{eq:motion} becomes
\begin{equation}\label{eq:2}
    \frac{1}{n^2(U,U)} = F\left(r\left(\frac{g(U)-f(U)}{2}\right)\right)f'(U)\,g'(U) \, .
\end{equation}
Using the known values of $g(V)$ for $0<V<L$ and $n(U,U)$ for $U>0$, we can determine $f(U)$ for $0<U<L$.

By iterating the previous steps, we can determine $f(U)$ and $g(V)$ for $U>-L$ and $V>0$, thus obtaining the motion of the rod: the compression factor can be computed at any point $(U,V)$ using equation \eqref{eq:motion}, and the trajectory of any particle in the rod, located at $\lambda=\lambda_0$ , is given by the curve
\begin{equation}
    \gamma(V) = \left(\frac{f(V-2\lambda_0)+g(V)}{2},\, r\left(\frac{g(V)-f(V-2\lambda_0)}{2}\right)\right), \qquad V>\lambda_0 \, ,
\end{equation}
where the first coordinate is the time $t$ and the second coordinate is the radius $r$.
%
%
\section{Energy}\label{section2.5}

Since $\frac{\partial}{\partial t}$ is a Killing vector field, there is a conserved energy for the motion of the rod, given by integrating on the level sets of $\tau$ the contraction of the energy-momentum tensor $T$, given in \eqref{eq:energymomentumtensor}, with the Killing vector field and the future-pointing unit normal:
\begin{equation}
E(\tau)=\int_0^L T\left(\frac{\partial}{\partial t}, n\frac{\partial}{\partial \tau}\right) \frac{d\lambda}{n} = \int_0^L T\left(\frac{\partial}{\partial t}, \frac{\partial}{\partial \tau}\right) d\lambda
\end{equation}
(see also \cite{Natario14, Natario18}). Using equations \eqref{eq:nullcoords}, \eqref{eq:nullcoords2} and \eqref{eq:nullcoords3} to express $\frac{\partial}{\partial t}$ as a linear combination of $\frac{\partial}{\partial \tau}$ and $\frac{\partial}{\partial \lambda}$, and then contracting with $T$ using \eqref{eq:energymomentumtensor}, we obtain
\begin{equation}\label{eq:energy}
E(\tau)= \int_0^L \frac{n^2(\tau,\lambda)+1}{4n^2(\tau,\lambda)} \left( \frac1{f'(\tau-\lambda)} + \frac1{g'(\tau+\lambda)} \right) \rho_0 \, d\lambda \, .
\end{equation}
The conservation of $E(\tau)$ will be used to double check the soundness of the numerical integrations to follow. 

It is also interesting to compute the (non-conserved) sum of the energies of the individual particles making up the rod, given by integrating on the level sets of $\tau$ the inner product of the Killing vector field with the unit tangent vector to their worldlines (which coincides with the future-pointing unit normal), wheighted by the relaxed density:
\begin{equation}\label{eq:mechanicalenergy}
K(\tau)=- \int_0^L g\left(\frac{\partial}{\partial t}, n\frac{\partial}{\partial \tau}\right) \rho_0 \, d\lambda = \int_0^L \frac{1}{2n(\tau,\lambda)} \left( \frac1{f'(\tau-\lambda)} + \frac1{g'(\tau+\lambda)} \right) \rho_0 \, d\lambda \, .
\end{equation}
The difference $E(\tau)-K(\tau)$ can be interpreted as the elastic energy of the rod. Monitoring the evolution of $K(\tau)$ will signal whether elastic energy is being converted to mechanical energy (meaning the combined kinetic and gravitational potential energy) or vice-versa.
%
%
\section{Falling into a Schwarzschild black hole}\label{section3}

We will now study the radial motion of an elastic rod in the Schwarzschild spacetime. The metric of the $2$-dimensional totally geodesic submanifold corresponding to a fixed radial direction is
\begin{equation}
    ds^2=-\left(1-\frac{1}{r}\right)dt^2+\left(1-\frac{1}{r}\right)^{-1}dr^2 \, ,
\end{equation}
where, by simplicity, we take the Schwarzschild radius to be our length unit, that is, $2M = 1$. In this case we have $F(r)=G(r)=1-\frac{1}{r}$, and so
\begin{equation}
    x(r) = \int\left(1-\frac{1}{r}\right)^{-1}dr=r+\ln{(r-1)} \, ,
\end{equation}
frequently referred to as the tortoise coordinate. Note that we assume $r>1$, implying that the rod is totally outside the black hole.

We will concentrate on the strong field region near the black hole, and present three illustrative examples. Specifically, we will discuss the motion of a rod of undeformed lenght $L=1$ (thus equal to the black hole radius) and relaxed density $\rho_0=1$,\footnote{The only role played by the value of $\rho_0$ is to fix the scale of the energies in equations \eqref{eq:energy} and \eqref{eq:mechanicalenergy}.} starting at rest, for three different choices of the initial compression factor: completely relaxed (for comparison purposes), and two cases where the two halves of the rod are uniformly compressed or streched by different amounts. Other examples, such as the uniformly compressed or stretched rod, yield comparable but less dramatic results. It is, of course, impossible to fully explore the infinite-dimensional space of initial conditions for the rod; the examples that we present (besides the natural initially relaxed configuration) correspond to some of the most extreme examples we found in our simulations, while keeping the initial conditions reasonably sensible.
%
%
\subsection{Initially relaxed free-falling rod}

\begin{figure}[H]
    \centering
    \includegraphics[width=0.5\textwidth]{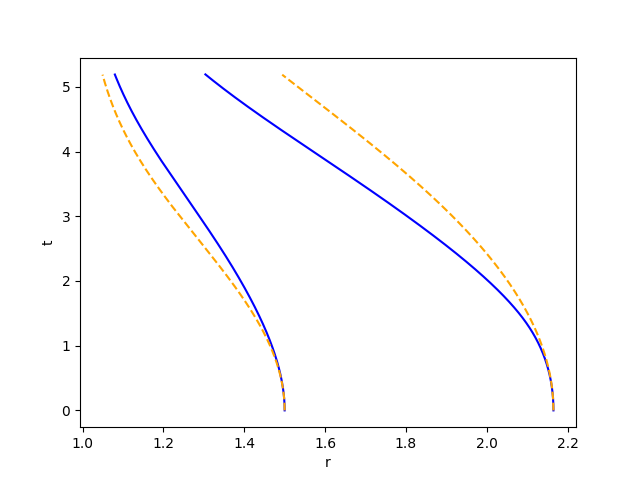}
    \caption{Initially relaxed free-falling rod in the Schwarzschild spacetime. The worldlines of the left and right endpoints are the blue curves. The geodesics corresponding to particles starting at rest from the same positions are the orange dashed curves.}
    \label{fig:image4}
\end{figure}

As a first example, we will explore the scenario in which the rod is initially completely relaxed and falls freely into the black hole. Mathematically, this corresponds to setting $n=1$ at $\tau=0$. Since equation \eqref{eq:motion} cannot be solved explicitly, one needs to resort to numerical integration. We use the Runge-Kutta method, repeating the integration for successively smaller time steps in order to check the convergence of the solution. The results are depicted in Figure~\ref{fig:image4}, where the geodesics representing test particles starting at rest from the same points as the rod's endpoints are also presented for comparison. We can see that the endpoint nearer the black hole falls slower than the particle starting from the same position, whereas the opposite endpoint falls faster. This is due to the rod's elastic nature, which exerts an elastic force on its endpoints, forcing them to move slightly inwards.

\begin{figure}[H]
    \centering
    \includegraphics[width=0.5\textwidth]{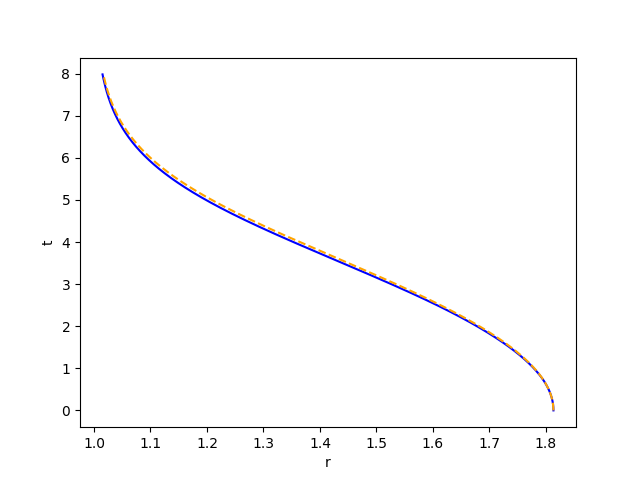}
    \caption{Worldline of the midpoint of the rod, initially relaxed and falling freely in the Schwarzschild spacetime. The worldline of the midpoint is the blue curve. The geodesic corresponding to a particle starting at rest from the same position is the orange dashed curve.}
    \label{fig:image11}
\end{figure}

We can also plot the trajectory of the midpoint of the rod, that is, the point corresponding to $\lambda = L/2$, and compare it with the geodesic corresponding to the free-falling test particle starting at rest from the same position. As shown in Figure~\ref{fig:image11}, the midpoint's worldline is significantly closer to the test particle's geodesic than the endpoints, which may be due to the fact that the action of the elastic forces is in both directions in this case, rather than just in one.

\begin{figure}[H]
    \centering
    \includegraphics[width=0.5\textwidth]{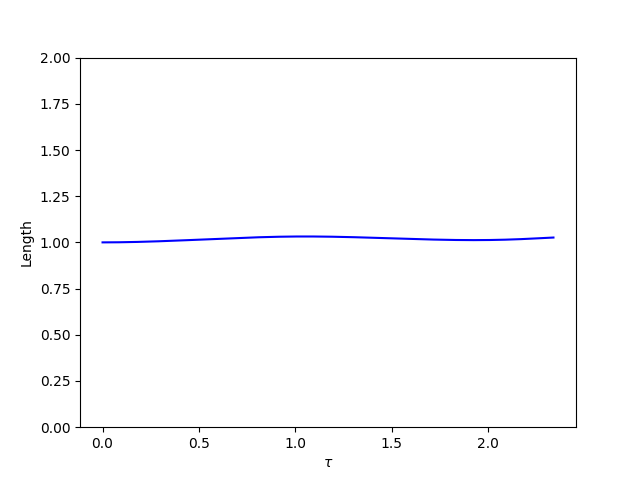}
    \caption{Total physical length (as measured by the comoving observers) of an initially relaxed rod falling freely in the Schwarzschild spacetime.}
    \label{fig:imagex}
\end{figure}

Finally, it is also instructive to plot the total physical length of the rod (as measured by the comoving observers), given by
\begin{equation}\label{eq:physicallength}
L(\tau)=\int_0^L \frac1{n(\tau,\lambda)} d\lambda \, .
\end{equation}
This is done in Figure~\ref{fig:imagex}, where it can be seen that the physical length is roughly constant throughout the motion.

As an aside, we can study the initial behavior of the compression factor --- in particular, whether it is initially increasing or decreasing. To do that, we use equations~\eqref{eq:motion} and \eqref{eq:1initial} to calculate the derivative of the quantity $\bar{n} \coloneqq 1/n^2$ with respect to $\tau$ at $\tau=0$, that is, at the points satisfying $U=-V$, for $0<V<L$:
\begin{equation}
    \frac{\partial\bar{n}}{\partial \tau}(-V,V)=\frac{\partial \bar{n}}{\partial U}(-V,V)+\frac{\partial \bar{n}}{\partial V}(-V,V) \, , \qquad 0<V<L \, .
\end{equation}
Using~\eqref{eq:motion}, we obtain:
\begin{align}
        \frac{\partial\bar{n}}{\partial\tau}(U,V) = & \,\, \frac{1}{2}F'\left(r\left(\frac{g(V)-f(U)}{2}\right)\right)r'\left(\frac{g(V)-f(U)}{2}\right)\left(f'(U)\left(g'(V)\right)^2-\left(f'(U)\right)^2g'(V)\right) \nonumber \\
        &+ F\left(r\left(\frac{g(V)-f(U)}{2}\right)\right)\left(f''(U)g'(V)+f'(U)g''(V)\right) \, .
\end{align}
Equation~\eqref{eq:derivs} implies that this derivative vanishes when evaluated at points of the form $(-V,V)$, with $0<V<L$. Therefore, we proceed to calculate the second derivative of $\bar{n}$ with respect to $\tau$ at $\tau=0$. By making use of equation \eqref{eq:derivs} to relate the derivatives of $f$ and $g$, we obtain:
\begin{align}
        \frac{\partial^2 \bar{n}}{\partial\tau^2}(-V,V) = & \,\, F'(r(g(V)))r'(g(V))\left(g'(V)\right)^2 g''(V)-2F(r(g(V)))\left(g''(V)\right)^2 \nonumber \\
        &+2F(r(g(V)))g'(V)g'''(V) \, .
    \end{align}
We can calculate the derivatives of the function $g$ from equation \eqref{eq:gderiv}. After a lengthy computation, we arrive at
\begin{equation}
    \frac{\partial^2 \bar{n}}{\partial\tau^2}(-V,V) = \frac{2}{\left(r(g(V))\right)^3} \, .
\end{equation}
Therefore, the second derivative of $\bar{n}$ is positive, indicating that the rod starts out by stretching. This can be understood as the effect of the tidal forces produced by the black hole.
%
%

\subsection{Gliding motion in the Schwarzschild spacetime}

\begin{figure}[H]
    \centering
    \subfloat[][]{
    \includegraphics[width=0.47\textwidth]{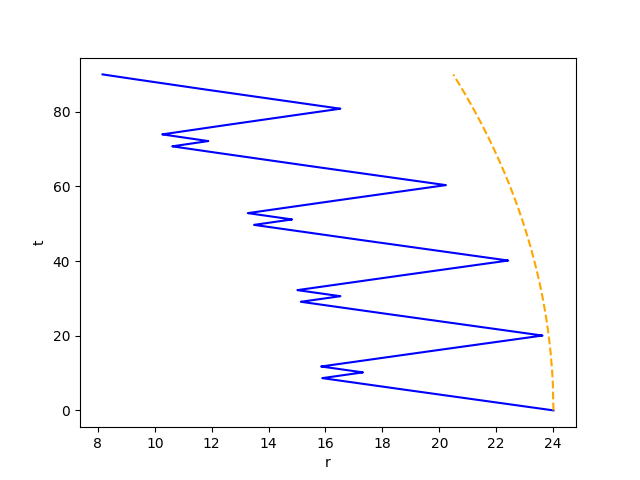}
    }
    \subfloat[][]{
    \includegraphics[width=0.47\textwidth]{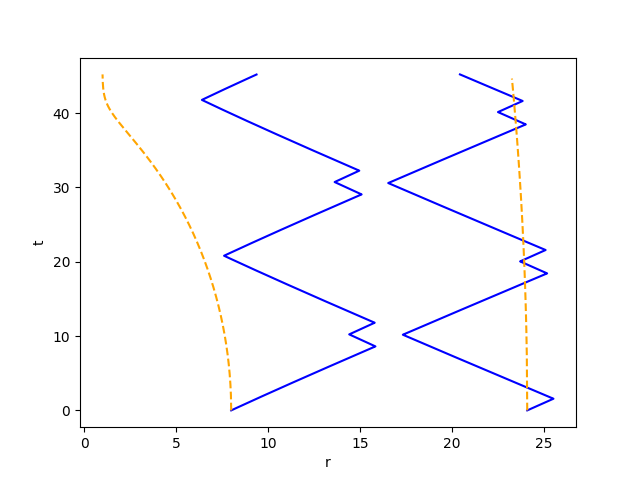}
    }
    \caption{Worldlines of (a) the midpoint, and (b) the endpoints of the initially compressed free-falling rod in the Schwarzschild spacetime, with initial pa\-ra\-me\-ters $n_1=0.03$ and $n_2=6$. The worldlines of points in the rod are the blue curves, while the geodesics corresponding to particles starting at rest from the same points are the orange dashed curves.}
    \label{fig:glider1}
\end{figure}

Finally, we will study the motion of a rod whose initial configuration is given by\footnote{We are calling the motions resulting from these initial conditions ``gliding motions" because, as we will see, shape of the rod will vary substantially and in a roughly periodic way, resulting in large deviations with respect to geodesic motion.}
\begin{equation}
    n(-V,V) = n_1(1-H(V, \varepsilon)) + n_2 H(V,\varepsilon) \, , \qquad 0<V<L \, ,
\end{equation}
where $n_1$ and $n_2$ are two positive constants and $H$ is the function
\begin{equation}
    H(V,\varepsilon) = \frac{1}{2}+\frac{1}{2}\tanh\left(\frac{V-\frac{L}{2}}{\varepsilon}\right) \, .
\end{equation}
Note that if we choose $\varepsilon>0$ sufficiently small then $H$ is a smooth approximation of a Heaviside step function that jumps from $0$ to $1$ at $V=L/2$, the midpoint of the rod. In our simulation we considered $\varepsilon = 10^{-3}$. With this definition, the initial configuration corresponds to a rod whose compression factor is $n_1$ on the left half of the rod and $n_2$ on the right half. For instance, by choosing the parameters $n_1=0.03$ and $n_2=6$ we obtain the result depicted in Figure~\ref{fig:glider1}.

For this choice of parameters, we can see that the rod's midpoint falls significantly faster than a free-falling particle starting at rest from the same point. 

\begin{figure}[H]
    \centering
    \subfloat[][]{
    \includegraphics[width=0.47\textwidth]{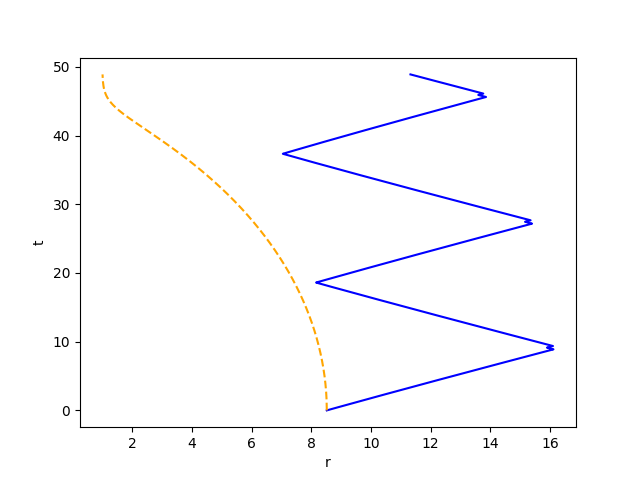}
    }
    \subfloat[][]{
    \includegraphics[width=0.47\textwidth]{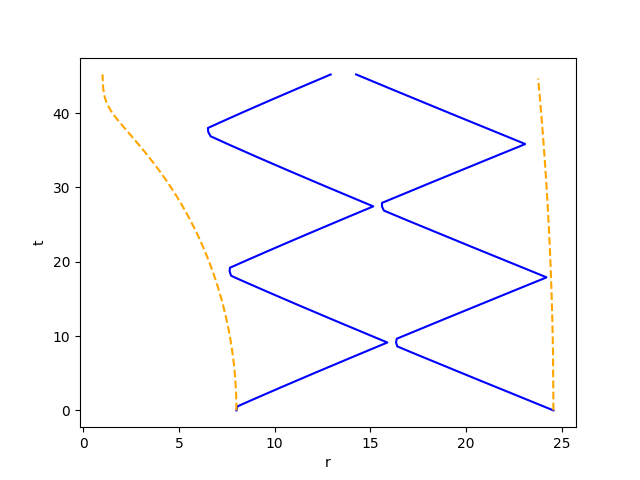}
    }
    \caption{Worldlines of (a) the midpoint, and (b) the endpoints of the initially compressed free-falling rod in the Schwarzschild spacetime, with initial pa\-ra\-me\-ters $n_1=0.9$ and $n_2=0.03$. The worldlines of points in the rod are the blue curves, while the geodesics corresponding to particles starting at rest from the same points are the orange dashed curves.}
    \label{fig:glider2}
\end{figure}

On the other hand, if we choose the parameters appropriately then we can set up an initial configuration such that the midpoint of the rod falls much slower than the particle. For example, the case with $n_1=0.9$ and $n_2 = 0.03$ is shown in Figure~\ref{fig:glider2}, where we can clearly see that the rod's midpoint falls slower than the free-falling particle starting at rest from the same initial position: not only does it not cross the geodesic, but the distance between the midpoint and the particle after each complete oscillation increases over time.

\begin{figure}[H]
    \centering
    \subfloat[][]{
    \includegraphics[width=0.47\textwidth]{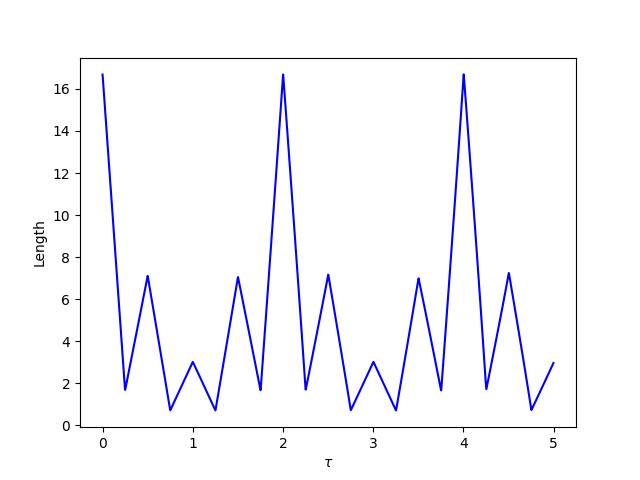}
    }
    \subfloat[][]{
    \includegraphics[width=0.47\textwidth]{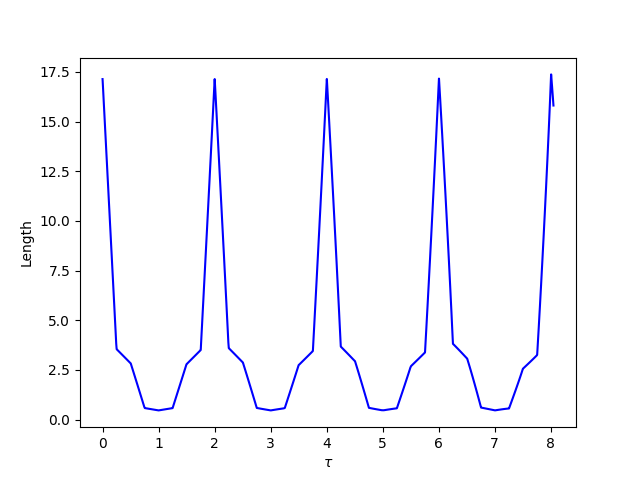}
    }
    \caption{Total physical length of the rods (as measured by the comoving observers) for (a) $n_1=0.03$ and $n_2=6$, and (b) $n_1 = 0.9$ and $n_2=0.03$. The rod seemingly oscillates periodically between states where its physical length is much larger that $L$ and states where its physical length is roughly equal to $L$.}
    \label{fig:lengths}
\end{figure}

Notice that these choices of parameters are quite extreme, corresponding to initial configurations where half the rod is drastically stretched. This can be seen in the plots of the total physical length of the rod (as measured by the comoving observers), shown in Figure~\ref{fig:lengths}, where it is apparent that the rod oscillates periodically between states where its physical length is much larger that $L$ and states where its physical length is roughly equal to $L$.

\begin{figure}[H]
    \centering
    \subfloat[][]{
    \includegraphics[width=0.47\textwidth]{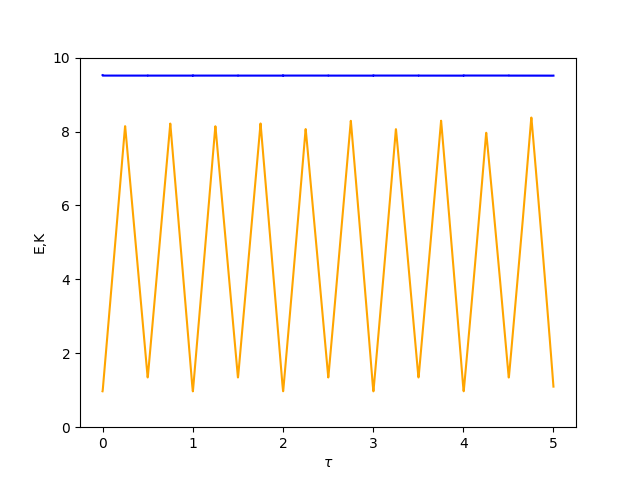}
    }
    \subfloat[][]{
    \includegraphics[width=0.47\textwidth]{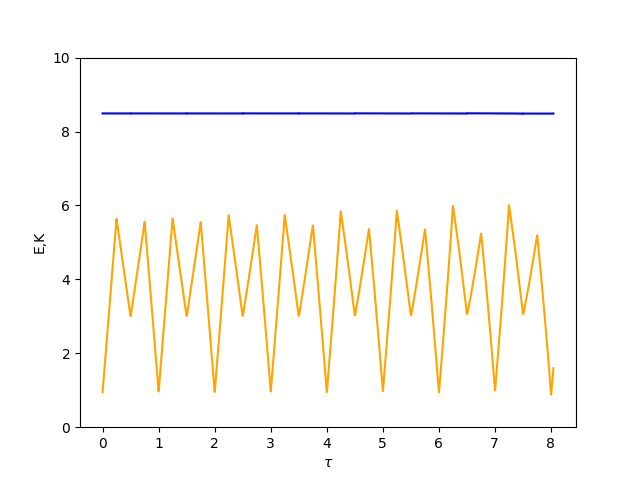}
    }
    \caption{Quantities $E$ (blue curve) and $K$ (orange curve) as a function of $\tau$ for (a) $n_1=0.03$ and $n_2=6$, and (b) $n_1 = 0.9$ and $n_2=0.03$. }
    \label{fig:energy}
\end{figure}

In order to check energy conservation and other energy considerations, the quantities $E$ and $K$ are presented as a function of $\tau$ in Figure~\ref{fig:energy} for the two cases above (the integrals were computed using Simpson's rule). We can see that in both cases the total energy of the rod, $E$, is constant, corroborating the soundness of the numerical integration of the equations of motion. The results depicted in Figure~\ref{fig:energy} also show that there is a periodic conversion of elastic energy to mechanical energy and vice-versa. However, there does not seem to be a net transfer from one form of energy to the other on average; the faster or slower falling motion of the rod when compared to geodesics cannot be attributed to an accumulation or depletion of elastic energy.

%
%
\section{Conclusion}\label{section4}

In this paper we considered, for the first time, the free-falling motion of an extended elastic body in the Schwarzschild gravitational field. For an initially relaxed rod, we saw that the motion of the midpoint is very close to pure geodesic motion, although the endpoints do deviate from the corresponding geodesics (as expected for an elastic body). What is interesting about the other two examples we presented, however, is the fact that the speed of the rod's midpoint can be significantly affected by the initial compression, when compared to pure geodesic motion.

In all these examples, only the initial compression profile can be chosen. Naturally, an interesting question would be to study the possibility of controlling the motion by some internal mechanism that would allow for changing the rod's shape as it falls. This would necessitate a more complicated mathematical model, in which energy flow along the rod would have to be included.

This $(1+1)$-dimensional formalism can only be used to study radial fall. A more complete description of the free-falling motion of a relativistic elastic body on the Schwarszchild field would require the full $(1+3)$-dimensional theory of relativistic elasticity, whose implementation can be considerably more difficult.
%
%
%
\section*{Acknowledgements}
This work was partially supported by FCT/Portugal through CAMGSD, IST-ID (projects UIDB/04459/2020 and UIDP/04459/2020) and project NoDES
(PTDC/ MAT-PUR/1788/2020), and also by the H2020-MSCA-2022-SE project EinsteinWaves, GA no.~101131233. LM gratefully acknowledges the Calouste Gulbenkian Foundation for the scholarship program \emph{Novos Talentos em Matem\' atica}.
%
%

\end{document}